\documentclass{scrartcl}

\usepackage[utf8]{inputenc}

\usepackage{float}
\usepackage{subfig}
\usepackage{graphicx}
\usepackage{amsmath}
\usepackage{hyperref} 
\usepackage{natbib}
\usepackage{amssymb}
\usepackage{algorithm}
\usepackage{algorithmic}
\usepackage{authblk}

\title{Explaining heatwaves with machine learning}

\author[1]{Sebastian Buschow}
\author[2,3]{Jan Keller}
\author[1,2]{Sabrina Wahl}

\affil[1]{Institute of Geosciences, University of Bonn}
\affil[2]{Hans-Ertel-Centre for Weather Research}
\affil[3]{Deutscher Wetterdienst}

\begin{document}

\maketitle

\begin{abstract}
    Heatwaves are known to arise from the interplay between large-scale climate variability, synoptic weather patterns and regional to local scale surface processes. While recent research has made important progress for each individual contributing factor, ways to properly incorporate multiple or all of them in a unified analysis are still lacking. 
    In this study, we consider a wide range of possible predictor variables from the ERA5 reanalysis, and ask, how much information on heatwave occurrence in Europe \textit{can be learned} from each of them. 
    To simplify the problem, we first adapt the recently developed logistic principal component analysis to the task of compressing large binary heatwave fields to a small number of interpretable principal components. The relationships between heatwaves and various climate variables can then be learned by a neural network. Starting from the simple notion that the importance of a variable is given by its impact on the performance of our statistical model, we arrive naturally at the definition of Shapley values. Classic results of game theory show that this is the only fair way of distributing the overall success of a model among its inputs. 
    With this approach, we find a non-linear model that explains 70\,\% of reduced heatwave variability, 27\,\% of which are due to upper level geopotential while top level soil moisture contributes 15\,\% of the overall score. In addition, Shapley interaction values enable us to quantify overlapping information and positive synergies between all pairs of predictors. 

\end{abstract}


\newpage
\section{Introduction}

Summer heat extremes are perhaps the most acutely alarming symptom of a warming global climate. Recent studies have identified Europe as a regional hot spot of such events \citep{russo2015,rousi2022}. Here, as in many other regions of the world, the hottest and most dangerous extremes occur not in isolation but as uninterrupted streaks of several days, called \textit{heat waves}. To arrive at a comprehensive explanation of the causes and characteristics of heat waves, several interacting parts of the climate system have to be taken into account. \cite{perkins2015} list synoptic weather systems, surface and soil moisture interactions and large scale climate variability as the three main categories of physical drivers. 
Motivated in part by the disastrous European heat waves of 2003 and 2010, much effort has been directed at understanding each of these. For example, the role of the planetary circulation has been investigated by \cite{kornhuber2019} and \cite{rousi2022}. Recent studies related to synoptic patterns, especially atmospheric blocking, include \cite{brunner2018}, \cite{schaller2018} and \cite{kautz2022}. \cite{miralles2019} give a current overview of progress made on land-atmosphere feedbacks. Finally, some attention has been paid to the connection between heat waves and anomalous states of the Oceanic circulation \citep{duchez2016,qasmi2021}.

Advances have thus been made in each category of heatwave drivers individually. Their relative importance and possible interactions in different regions, on the other hand, is considered by \cite{perkins2015} as one of the main open research directions. \cite{grotjahn2016} list composites, regression, empirical orthogonal functions and clustering as the most widely used statistical tools to identify meteorological patterns related to heat waves. In their classic forms, none of these techniques are well suited to the analysis of multiple correlated and interacting variables at once. In addition, many typical approaches explicitly or implicitly assume normally distributed data and linear relationships. 

In this study, we follow a more recent methodological trend \citep{mcgovern2019,toms2020,tesch2022}: Instead of employing statistical tools of insufficient complexity or numerical models with intractable cost, we first emulate the observed processes by a machine learning model. In a second step, we attempt to explain the success of that model.
Multiple linear regression, as used by \cite{suarez2020}, can be seen as a special case of this idea since this model explains itself via its regression coefficients. It is not, however, ideal for the study of non-linearities and interactions. 
\cite{felsche2021} and \cite{vanStraaten2022} employ more powerful regression models (neural nets and random forests) to predict droughts and heatwaves, respectively. The coefficients of those models are generally not human-interpretable. These studies therefore rely on the externally applied explanation framework of \cite{lundberg2017}, which computes estimates of the game theoretic Shapley values \citep{shapley52} to decompose the model output into additive contributions from all the inputs. 
In this study, we use Shapley values in a subtly different way: instead of explaining what a single model \textit{has learned}, we ask how much \textit{can be learned} from a range of possible predictors. Following \cite{hausken2001}, we furthermore extend these explanations to explicitly quantify interactions: how does the amount of relevant information from predictor $i$ change if we also include predictor $j$? 

The data from the ERA5 reanalysis \citep{hersbach2020} used throughout this study is summarized in section \ref{sec:data}. 
A suitable model to reconstruct heatwaves from other climate variables is developed in section \ref{sec:nnet}. Here, we show how binary fields can be compactly represented by the recently developed logistic PCA of \cite{landgraf2020}. We put this method in the context of neural nets, allowing us to efficiently apply and adapt it to our needs. The compressed heatwaves are then reproduced by simple neural networks. Section \ref{sub:expl} introduces Shapley values and interactions as the natural way of quantifying the role of individual predictors. The methods are subsequently applied to the heatwave problem (section \ref{sec:res}). In section \ref{sec:disc}, we discuss in detail what kinds of statement can and cannot be made based on our analysis. Lastly, we summarize our results and discuss directions for future research (section \ref{sec:outro}).


\section{Data}\label{sec:data}
For this study, we rely on the ERA5 reanalysis \citep{hersbach2020}, which provides a spatio-temporally consistent estimate of the state of the global climate system. We include the preliminary back-extension, giving us an uninterrupted time-series of 71 years from 1950 to 2020. For Europe in the latter half of the data-set (1979 onwards), \cite{keller2021} have shown that heat related climate indices are adequately represented. The data is provided on a $0.25^\circ$ geographical grid with 137 vertical pressure levels at hourly time steps. For most variables, we compute daily averages from all hourly time-steps. Our heatwave definition relies on daily maxima instead; for geopotential and wind components on multiple vertical levels, we retrieved the 13:00\,UTC (average time of maximum temperatures) fields to save time and disk space. 
All fields are bilinearly interpolated to a rotated $0.22^\circ$ grid of the European CORDEX domain with $212\times 206$ grid points (see figure \ref{fig:HW}). 

\subsection{Heatwave definition}
\begin{figure}
    \centering
    \includegraphics[width=.66\textwidth]{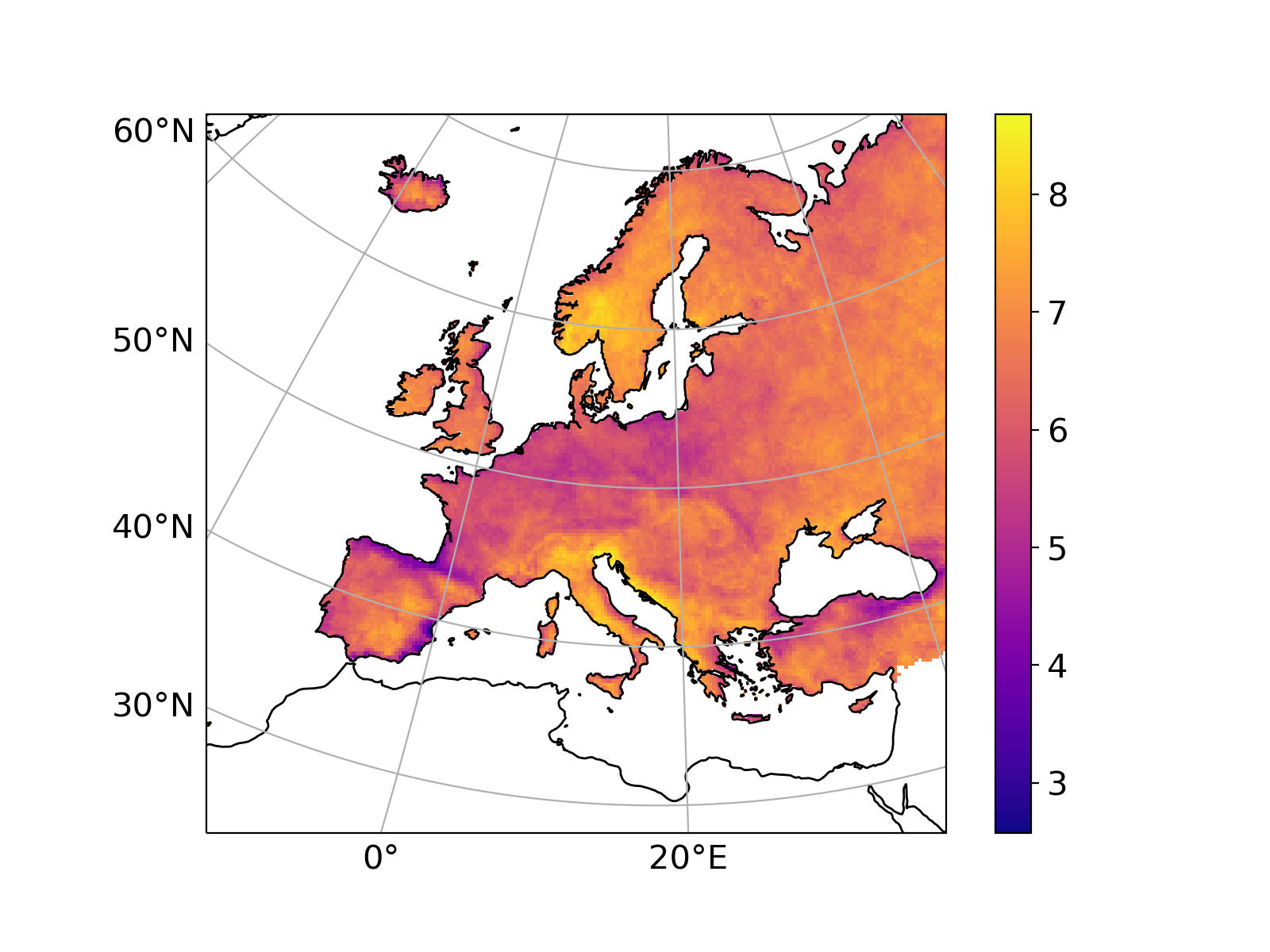}
    \caption{Percentage of summer days in ERA5 (JJA 1950-2020) that are part of a heatwave (10\,\% of days are hot by definition, heat waves are streaks of 3 or more hot days in a row).}
    \label{fig:HW}
\end{figure}

A large number of heatwave definitions exists in the literature, with no generally agreed upon standard \citep{grotjahn2016}. For our purposes we select a relatively simple approach based on daily maximum temperatures alone, which allows heatwaves to occur in any location at any time. To this end, we follow \cite{lavaysse2018} and define a day as \textit{hot} if the daily maximum temperature exceeds the climatological 90\,\% quantile for the current calendar day. The quantile thresholds are computed over a sliding 11 day window over the whole time series.\footnote{The threshold for June 6 is estimated from June 1-11 1950-2020 and so on.} A hot day is a \textit{heat wave} day if it is part of a streak four days, three or four of which are hot. By allowing for one gap day, we slightly increase the overall heatwave frequency and reduce sensitivity to day-to-day fluctuations in the temperature and the threshold. Thresholds, hot days and heat waves were computed for each European land grid point (colored are in figure \ref{fig:HW}) over the entire ERA5 time series; in this study we consider only the results for the summer months (JJA).\par

Figure \ref{fig:HW} reveals that, despite the local percentile thresholds, there is considerable spatial variability in the heatwave occurrence: while the frequency of hot days is fixed at 10\,\% everywhere, the percentage of these hot days that occur during a heat wave varies from around 50\,\% in central Europe up to 80\,\% in Scandinavia, Italy and along the coast of the Adriatic sea. Furthermore, heatwaves in the latter regions last for an average of five to six days compared to less than four days in Poland (not shown).

\subsection{Predictor variables}\label{sub:pred}

Considering the vast number of variables provided by ERA5, some pre-selection of relevant potential predictors is inevitable. If we assume that the dry synoptic dynamics related to European heat waves are approximately geostrophic, all information is contained in the 3D fields of geopotential. At the very least, two pressure levels are needed to describe the barotropic and baroclinic components of the system, the classic choices being 500 and 1000\,hPa. To explicitly include information from the jet stream along the lines of \cite{rousi2022}, we add the zonal wind in 200\,hPa as well.
Given the known importance of surface feedbacks, soil moisture and surface evaporation are two obvious additional candidate predictors. To round out the representation of the water cycle, we further include the total precipitation. Solar radiation, determined mainly by cloud cover, is another obvious candidate. 
Including temperature itself directly would be unhelpful in this context since we know its relationship with heatwaves by definition. Temperature advection, however, is added to the pool of model inputs. 
All variables included in this study are listed in table \ref{tab:predictors}. 

\begin{table}
    \centering
    \begin{tabular}{llll}
         name & height level& abbreviation  & note \\\hline
         zonal wind speed & 200\,hPa & u0200 & \\
         geopotential & 500\,hPa, 1000\,hPa  & z0500, z1000 & \\
         daily mean temperature advection & 1000\,hPa & adv1000m & standardized \\
         surface net solar radiation & surface & ssr & \\
         daily precipitation & surface & precip & standardized \\
         evaporation & surface & e & land only \\
         volumetric soil water & soil layer 1, 2 & VSW1, VSW2 & land only\\
         \hline 
    \end{tabular}
    \caption{List of predictor variables and their abbreviations used throughout this manuscript.}\label{tab:predictors}
\end{table}

\section{Methods}

\subsection{Neural networks}\label{sec:nnet}
We are interested in the relationships between heatwaves and various other climate variables. Mathematically, we study functions $f$ that map certain input variables $\Vec{x}\in\mathbb{R}^p$ to binary heat wave fields $\Vec{y}\in\mathbb{R}^q$. All of the statistical tools we use to simplify this problem, and approximate the real, complex relationships, can be understood as \textit{neural networks}. We consider the simple class of neural networks, given by functions $f:\mathbb{R}^p\to\mathbb{R}^q$ of the form
\begin{equation}
    f(\Vec{x}) = \Vec{x}^{(N)} \, \text{, with}\hspace{10pt} 
    \Vec{x}^{(i)} = \begin{cases}
    \Vec{x}\,,\hspace{5pt}\text{for }i=0\\
    \phi^{(i)}( \mathbf{W}^{(i)} \Vec{x}^{(i-1)} + \Vec{b}^{(i)} )\,\hspace{5pt}\text{otherwise.}
    \end{cases}
\end{equation}
The input $\Vec{x}$ is thus transformed by alternating application of linear transformations $\mathbf{W}\Vec{x} + \Vec{b}$  and element-wise \textit{activation functions} $\phi$. $\Vec{x}^{(0)}=\Vec{x}$ and $\Vec{x}^{(N)}=f(\Vec{x})$ are called the input and output layer, respectively. For $N>1$, the function additionally computes $N-1$ intermediate states $\Vec{x}^{(i)}\in\mathbb{R}^{r(i)}, i=1,2,...,N-1$, which are called \textit{hidden layers} of the neural network. The number and sizes $r(i)$ of the hidden layers determine the complexity of the model. It can be shown that a neural network with one hidden layer and an appropriate non-linear activation function  can approximate any function of interest, provided that the size $r(1)$ of the hidden layer is large enough. To obtain a neural network that performs some task of interest, the weight matrices $\mathbf{W}^{(i)}$ and bias vectors $\Vec{b}^{(i)}$ are iteratively adapted to minimize some loss function $J$. In an machine learning context, the iterations are referred to as \textit{epochs} and the process is called \textit{training} or \textit{learning}. Deeper (more layers) and wider (larger layers) networks can potentially learn more complicated relationships, but are more difficult to optimise and more prone to over-fitting.\\

\paragraph{Regression}
As our first application, we consider the task of predicting some output $\Vec{y}\in\mathbb{R}^q$ given an input $\Vec{x}\in\mathbb{R}^p$ such as to minimize the mean squared error $J=|f(\Vec{x})-\Vec{y}|^2$. If we set $N=1$ (no hidden layer) and $\phi_1(\Vec{x})=\Vec{x}$ (identity function as activation), our neural network model is simply a (multivariate) linear regression.\\
Next, we are interested in the case of binary target vectors $\Vec{y}\in\{0,1\}^q$;  $\Vec{x}$ contains real numbers as before. In this case, the elements of $\Vec{y}$ represent the occurrence (1) or non occurrence (0) of an event  at some location and we want $f$ to have outputs between 0 and 1 which can be interpreted as the probability of that event. The natural loss function in this setting is the negative Bernoulli log-likelihood
\begin{equation}
    J_B(\Vec{x},\Vec{y}) = -\sum_{i=1}^q\log\left( [f_i(\Vec{x})]^{y_i} \cdot [1-f_i(\Vec{x})]^{1-y_i} \right)\,,\label{eq:BCE}
\end{equation}
which is also known as (binary) cross entropy. If we optimize a network with this loss function, set $N=1$ and use the \textit{logstic} (or \textit{sigmoid}) activation function
\begin{equation}
  \phi_1 = \sigma(x) = \frac{1}{1+e^{-x}}\in[0,1]\,,\label{eq:sigmoid}
\end{equation}
thereby guaranteeing output between 0 and 1, we obtain the well-known \textit{logistic regression}.\\ 
 
\paragraph{Compression}
In principle, we could model the relationship between fields of geopotential and our binary heat wave fields by a single logistic regression model. This would lead to a weight matrix $\mathbf{W}^{(1)}\in\mathbb{R}^{p\times p}$ where $p$ is the number of grid points. A model with so many free parameters is difficult to train, evaluate and interpret and thus unsuitable to our task. As mentioned above, our strategy is instead to reduce the dimensions of both $\Vec{x}$ and $\Vec{y}$ before attempting to model their relationship.\\ 
To reduce the dimensions of $\Vec{x}$, we can employ a neural network with one hidden layer to optimize the mean squared error $J(\Vec{x}) = |f(\Vec{x})-\Vec{x}|^2$. In contrast to the regression problems above, the output of the network is thus optimized to approximate the input $\Vec{x}$ itself. When the size of the hidden layer $r(1)$ is much smaller than $p$, the network will attempt to project $\Vec{x}$ to some lower dimensional subspace, from which the original $\Vec{x}$ can be reconstructed as best as possible in terms of the MSE. In other words, the hidden state $\Vec{x}^{(1)}$ is chosen such that it contains as much of the information from the full $\Vec{x}$ as possible. The part of the network that maps $\Vec{x}$ to $\Vec{x}^{(1)}$ is called the \textit{encoder}, the part that maps back is the \textit{decoder} and the overall model is referred to as an \textit{autoencoder}. If we use the identity for the link functions $\phi_1,\phi_2$, the resulting linear autoencoder performs a \textit{principal component analysis} (PCA) in the sense that $\Vec{x}^{(1)}$ is in the same subspace as the first $r(1)$ principal components. In general, the rows and columns of $\mathbf{W}^{(1,2)}$ are not equal to the basis vectors of the PCA; however, these vectors may be recovered by adding an $L_2$ penalty term (known as \textit{weight decay} in a neural network context) to the cost function \citep{kunin2019}. In a climate context, the basis vectors are known as empirical orthogonal functions (EOFs), the projection onto these vectors (values of the hidden layer) are the principal components (PCs). 

We could use the same method to obtain a reduced version of $\Vec{y}$ as well. Since, in our case, $\Vec{y}$ represents binary heat wave fields, we again replace the final activation function $\phi_2$ with $\sigma(x)$ (equation \ref{eq:sigmoid}) and use the cross entropy $J_B(\Vec{x},\Vec{x})$ (equation \ref{eq:BCE}) as our loss function. The resulting non-linear autoencoder is equivalent to the \textit{logistic principal component analysis} (lPCA) which was recently introduced by \cite{landgraf2020}.\\

\paragraph{Setup and implementation}
In summary, we approach the problem of mapping climate variables $\Vec{x}$ to binary heat waves $\Vec{y}$ by first reducing the dimensions of $\Vec{x}$ via a classic PCA (linear autoencoder) and $\Vec{y}$ via lPCA (simple non-linear autoencoder). In a second step, we train a neural network to predict the reduced heat waves $\Vec{y}^{(1)}$ from the reduced predictor variables $\Vec{x}^{(1)}$. We will compare a linear regression ($N=1$, identity activation) to a non-linear regression with $N=2$ and the non-linear activation function
\begin{equation}
    \text{ReLU}(x) = \max(x,0)\label{eq:relu}\,.
\end{equation}
This choice of $\phi_1$ is called a \textit{rectified linear unit} (ReLU), which explains the name ``activation function'': the $i$-th element of the hidden layer makes a non-zero contribution to the output if and only if $(\mathbf{W}\Vec{x})_i>-b_i$, otherwise it is \textit{inactive} and contributes nothing. All of the algorithms discussed in this section are summarized in table \ref{tab:mod_overview} in terms of their neural network equation and optimized loss function.\\

\begin{table}
    \centering
    \begin{tabular}{lll}
          model &  $f(\Vec{x})=$&  loss \\\hline
         linear regression & $\mathbf{W}_1\cdot\Vec{x}+\Vec{b}_1$ & $|f(\Vec{x})-\Vec{y}|^2$\\
         logistic regression & $\sigma( \mathbf{W}_1\cdot\Vec{x}+\Vec{b}_1)$ & $J_B(\Vec{x},\Vec{y})$\\
         non-linear regression & $\mathbf{W}_2\cdot\text{ReLU}( \mathbf{W}_1\cdot\Vec{x}+\Vec{b}_1)+\Vec{b}_2$& $|f(\Vec{x})-\Vec{y}|^2$\\
         PCA & $\mathbf{W}_2\cdot\mathbf{W}_1\cdot\Vec{x}$ & $|f(\Vec{x})-\Vec{x}|^2$\\
         lPCA & $\sigma(\mathbf{W}_2\cdot\mathbf{W}_1\cdot\Vec{x})$ &  $J_B(\Vec{x},\Vec{x}) + L_2(\mathbf{W}_i)$\\
         sparse lPCA & $\sigma(\mathbf{W}_2\cdot\mathbf{W}_1\cdot\Vec{x})$ &  $J_B(\Vec{x},\Vec{x}) + L_2(\mathbf{W}_i)+L_1(\mathbf{W}_i)$
    \end{tabular}
    \caption{Basic regression and compression algorithms, written as neural networks. For PCA and lPCA, a single set of basis vectors can be obtained by enforcing $\mathbf{W}_1=\mathbf{W}_2^T$.}
    \label{tab:mod_overview}
\end{table}

The observation that all of our algorithms are really neural networks enables us to implement and optimize them using widely available machine learning software. In particular, many of the necessary computations can efficiently be carried out on GPUs. Here, we rely on PyTorch \citep{pytorch} to implement both the linear and the non-linear regression, optimizing both through back-propagation using the \textit{Adam} algorithm \citep{kingma2014}.
For the standard PCA, we employ the usual method via partial eigen-decomposition of the covariance matrix (see for example \cite{wilks2011}), which is already sufficiently fast and guaranteed to yield the correct vectors in a single step and in order of their importance. 
For the lPCA, on the other hand, the implementation as a neural net is very appealing: the iterative majorization minimization algorithm proposed by \cite{landgraf2020} requires repeated multiplications between $p\times p$ matrices, where $p$ is the number of grid points. In our case, estimating the lPCA vectors in this way requires considerable computation time (roughly half an hour on 12 CPU cores) and cannot easily be scaled up to larger data-sets due to memory constraints. With Adam and a \textit{batch learning} approach, where manageable subsets of the data are processed in parallel, we can achieve almost identical results in seconds on a laptop GPU. A moderate amount of tuning was required to select an appropriate learning rate, strength of the $L_2$ penalty and initial scaling of the data.\par 

A further advantage of this implementation is the possibility to customize the data compression to our needs: to facilitate interpretation, we would like each basis vector to uniquely represent heatwaves in some localized part of the domain. With the neural network implementation of the lPCA, we can easily achieve this by adding an $L_1$ penalty term on the negative entries in the basis vectors. Like the Lasso approach known from regression analysis \citep{tibshirani1996}, this constraint tends to set large parts of the basis vectors to zero, encouraging a sparse representation. By applying it to the negative entries only, we arrive at basis vectors that are monopoles, such that positive values of the principal components are uniquely linked to the occurrence of a heatwave. We denote this as \textit{sparse logistic PCA} in the following.

\subsection{Shapley values}\label{sub:expl}

Having found models that can reproduce some part of the heatwave variability from a set of predictor variables $M=\{X_1,...,X_m\}$, we are interested in the contributions of each individual variable to the overall success of the model. Quantifying the success of a regression model with $m$ predictors by the coefficient of determination $R^2$, we seek an explanation of the form 
\begin{equation}
    R^2(M) = \varphi_0 + \sum_{i=1}^m\varphi_i\,,\label{eq:addex}
\end{equation}
where $\varphi_0$ is the performance of a model without any predictors and $\varphi_i$ represents the contribution from predictor variable $X_i$. Even for a linear model, it is not obvious how one should ``fairly'' distribute the total performance since the regression coefficients can become misleading when predictors are not independent \citep{lipovetsky2001}.

To arrive at a useful definition of $\varphi_i$, consider a model based on a subset of predictors $S\subset M$, not including $X_i$. Intuitively, the contribution of $X_i$ should be related to the difference between the previous skill $R^2(S)$ and the value $R^2(S\cup X_i)$ obtained when $X_i$ is added to the set of predictors. Based on this simple idea, we define the decomposition procedure given by algorithm \ref{alg:randomshap}.\\

\begin{algorithm*}
\caption{ Random decomposition of $R^2$ }\label{alg:randomshap}
    \begin{algorithmic}[1]
        \STATE train a model with no predictors, compute $\varphi_0=R^2_0=R^2(\emptyset)$
        \FOR{$k=1,...,m$} 
            \STATE randomly select a predictor $X_i$ that hasn't been chosen yet
            \STATE train a model with all predictors selected so far, compute its score $R^2_k$
            \STATE set $\varphi_i=R^2_k - R^2_{k-1}$
        \ENDFOR
    \end{algorithmic}
\end{algorithm*}

We can easily convince ourselves that the contributions $\varphi_i$ defined in this manner always sum up to the total $R^2$ (equation \ref{eq:addex}), since each individual $R^2_k$ is added and subtracted exactly once, except for the end result $R^2(M)$. Another desirable property is that a variable with zero impact on any model receives $\varphi_i=0$ every time.\\

It is clear that, in general, the output of algorithm \ref{alg:randomshap} depends on the random order in which the variables are added: if $X_i$ and $X_j$ contain a lot of overlapping information, the one that is added later elicits less of an improvement and gets a smaller $\varphi$. Conversely, $X_j$ may only be informative if we know $X_i$ already, such that adding it before $X_i$ yields little improvement. 
To take these effects into account, we simply set the $\varphi_i$ to the \textit{expectation values} of algorithm \ref{alg:randomshap}, i.e.,
\begin{equation}
    \varphi_i = E\left[R^2(S\cup X_i) - R^2(S)\right]\,,\label{eq:expshap}
\end{equation}
where $S$ is the set of variables drawn before $X_i$ and the expectation is taken over all possible random orders. The probability for any particular set $S$ of size $s$ to occur before $X_i$ is simply the probability $m^{-1}$ of drawing $X_i$ in the $s$-th position, multiplied by the probability of selecting one particular $S$ out of all possible combinations of $s$ predictors. With this probability
\begin{equation}
    Prob( S\text{ is drawn before }X_i ) := \gamma(s,m) = \frac{1}{m}\binom{m-1}{s}^{-1}\,,
\end{equation}
we can thus compute the expected contributions in general terms as
\begin{equation}
    \varphi_i(f,M) = \sum_{\text{all } S\subset M}  \gamma(s,m)\left[f(S\cup X_i) - f(S) \right]\,,\label{eq:shapley}
\end{equation}
where $f$ could be any function of a set of input variables $M$ and $\sum\varphi_i=f(M)$. This is the definition of the so-called \textit{Shapley values} of cooperative game theory, algorithm \ref{alg:randomshap} corresponds to the ``bargaining procedure'' described in section 6 of \cite{shapley52}. Their use for the analysis of regression models was suggested by \cite{lipovetsky2001} and recently popularized (in a slightly different setting) by \cite{lundberg2017}. With $f=R^2$, this is the exact result for the expectation value in equation \ref{eq:expshap}. Keeping in mind that the differences between $R^2$ with and without $X_i$ completely determine the outcome of algorithm \ref{alg:randomshap}, we recognize the following two additional properties of the expected $\varphi_i$:
\begin{itemize}
    \item \textit{symmetry}: if $X_i$ and $X_j$ have the same impact on all $S$ to which they could be added, then $\varphi_i=\varphi_j$.
    \item \textit{monotonicity}: if $f(S\cup X_i)-f(S)\geq g(S\cup X_i)-g(S)$ for all $S\subset M$,\\ then $\varphi_i(f,M)\geq\varphi_i(g,M)$.
\end{itemize}
In our context, the functions $f$ and $g$ might be the $R^2$ of two different models, or the same model trained on different data. 
Remarkably, it can be shown \citep{young1985} that the Shapley values defined by equation \ref{eq:shapley} are the only symmetric and monotonic decomposition rule. For our purpose of explaining model performance, missing symmetry would be unacceptable since identically behaving predictors should clearly be valued identically. Without monotonicity, we could not reasonably compare the impact of a particular predictor on different models or for different data-sets.

Exact computation of equation \ref{eq:shapley} requires us to train all $2^m$ possible models in order to obtain their $R^2$, which is only feasible for a small to moderate number of predictors. Fortunately, algorithm \ref{alg:randomshap} immediately suggests a way of estimating the $\varphi_i$ from a limited subset of all models: simply repeat the random decomposition $N$ times, requiring the training of fewer than $N\times m$ models, and average the results to get a sample approximation of the expected $\varphi_i$. A test of this approach is shown in appendix \ref{app:sample}.

\paragraph{Shapley interactions}
In addition to the overall contribution of each variable, we are interested in the interactions between different predictors: two variables with overlapping information will each have reduced importance when both are present. Conversely, some predictors may become more useful when another variable is also present. To quantify these negative and positive synergies, we follow \cite{hausken2001} and define the symmetric matrix of \textit{Shapley interactions} as 
\begin{equation}
\varphi_{i,j}(f,M) := \varphi_i( \varphi_j( f, \cdot ), M )\label{eq:hausken1}
    = \sum_{\text{all } S\subset M}  \gamma(s,m)\left[  \varphi_j( f, S \cup X_i ) - \varphi_j(f,S) \right]\,.
\end{equation}
Simply put, $\varphi_{i,j}$ is the average change in $X_i$'s contribution to the overall result when $X_j$ is added to the ``game'' (in our case the pool of possible predictors). Due to the properties of $\varphi$, the rows and columns of $\varphi_{i,j}$ add up to the original $\varphi_i$ and, consequently, all elements add up to $f(M)$. Negative entries $\varphi_{i,j}<0$ indicate that the presence of $X_j$ reduces $X_i$'s contribution compared to situations where $X_j$ is missing; $\varphi_{i,j}>0$ indicates positive synergy between the two variables. The diagonal elements $\varphi_{i,i}$ are the weighted mean of $X_i$'s Shapley values in all possible sub-games. \cite{hausken2001} interpret cases where  $\varphi_i>\varphi_{i,i}$ as cases in which an overall beneficial synergy is present for $X_i$, meaning that it is more valuable combined with other predictors than alone.


\section{Results}\label{sec:res}

\subsection{Logistic PCA}

\begin{figure*}
    \centering
    \includegraphics[width=\textwidth]{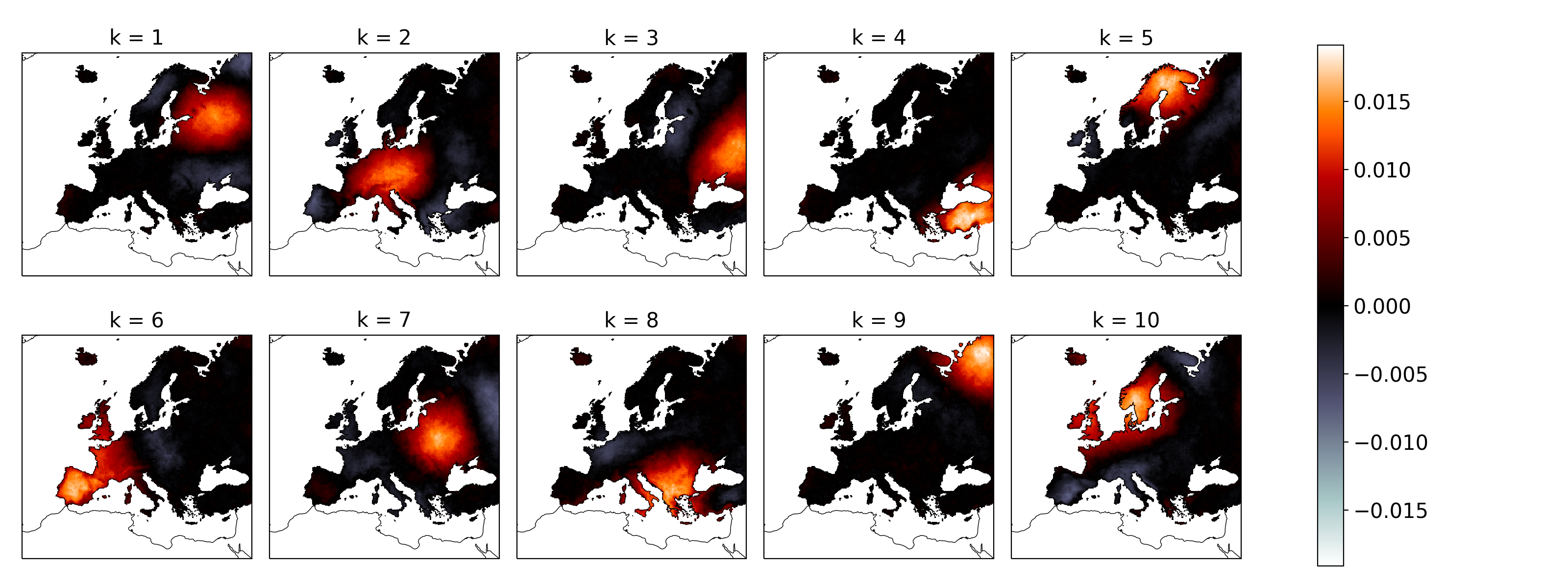}
    \caption{Ten sparse logistic EOFs of ERA5  heat wave fields (JJA 1950-2020).}
    \label{fig:logEOF}
\end{figure*}

We apply the sparse logistic PCA to the binary fields of ERA5 heatwaves to obtain a compressed representation in terms of ten logistic principal components. As intended, the corresponding basis vectors, shown in figure \ref{fig:logEOF}, are each localised to a specific part of the domain. Consequently, positive values of, for example, the tenth principal component are uniquely related to the occurrence of a heatwave in Southern Scandinavia. 

As in the case of a regular PCA, we can compute the fraction of variance explained by these ten principal components, resulting in a value close to 50\,\%. One should, however, keep in mind that the mean square error, and thus the variance, is not the natural quantity associated with Bernoulli distributed variables. An alternative would be the explained deviance, given by the improvement in Bernoulli likelihood over some reference model. However, this quantity is not as straightforward to interpret due to the somewhat arbitrary choice of reference. Regardless of our chosen metric (variance or deviance), the result of the lPCA is slightly degraded  by the additional $L_1$-penalty (relative difference up to 10\,\%, not shown).
For our purposes, this drawback is outweighed by the benefit of having well-localized basis functions, which allows for a straightforward interpretation of model results in different regions.

\subsection{Regression model performance}\label{sub:regr}
A linear and a non-linear neural network regression model are trained to predict the principal heat wave components corresponding to the ten basis vectors shown in figure \ref{fig:logEOF}. All predictors and predictands are standardized to zero mean and unit standard deviation. Standardizing the predictors tends to facilitate the learning process, while bringing all preditctands to the same scale encourages the model to pay equal attention to each region.
As discussed above, the linear neural network is identical to a classic linear regression. For the non-linear version, we use one hidden layer with a ReLU activation function. The two main free parameters in this set-up, i.e., the size of the hidden layer and the number of principal components from each predictor variable to include, were selected through a limited set of sensitivity experiments. All parameters are listed in table \ref{tab:nnetparams}. 

\begin{table}
\centering
        \begin{tabular}{lcc}
         & linear model (\textit{lm}) & non-linear model (\textit{nn})\\\hline
         \# input PCs & 20 per variable  & 20 per variable\\
         \# outputs & 10 & 10 \\
         hidden layer size & 0 & 40 \\
         activation $\phi_1$ & $\text{identity}(x)$ & $\text{ReLU}(x)$ \\
         loss & MSE & MSE \\
         learning rate & 0.05 & 0.001 \\
         dropout rate & 0 & 0.2 \\
         max \# epochs & 25 & 300\\
         patience & $\infty$ & 20 \\
         trained parameters & $\underset{20n\times 10}{\mathbf{W}_1}$, $\underset{10}{\Vec{b}_1}$ & $\underset{20n\times 40}{\mathbf{W}_1}$, $\underset{40}{\Vec{b}_1}$,\hspace{5pt}$\underset{40\times 10}{\mathbf{W}_2}$, $\underset{10}{\Vec{b}_2}$
    \end{tabular}
    \caption{Parameters of the two regression models. ``Patience'' denotes the number of epochs without improvements that are tolerated before training is stopped.}
    \label{tab:nnetparams}
\end{table}

Both models were trained and evaluated in a seven-fold cross-validation, where timestep $t$ is sorted into fold $Y_t\,\%\,7$. Here, $Y_t$ is the year of that time step and $\%$ indicates the modulo division. In this way, each fold contains years from all decades and no heat wave is split into multiple folds. To combat overfitting of the non-linear model, we employ a so-called \textit{dropout} step where $20\,\%$ of the neurons in the hidden layer are randomly deactivated during the optimization, effectively preventing the model from simply ``remembering'' already seen data points and incentivizing it to learn generalizable connections instead. Secondly, we add an early stopping criterion wherein the training is stopped if the best loss was observed more than $p$ epochs ago. To this end, the last 10\,\% of the training data are not used for optimization, giving us an independent test set on which to check the model improvement after each epoch. For the linear model, both of these measures can be deactivated and the learning procedure can be sped up without relevant loss of performance.\\

\begin{figure}
    \centering
    \includegraphics[width=.66\textwidth]{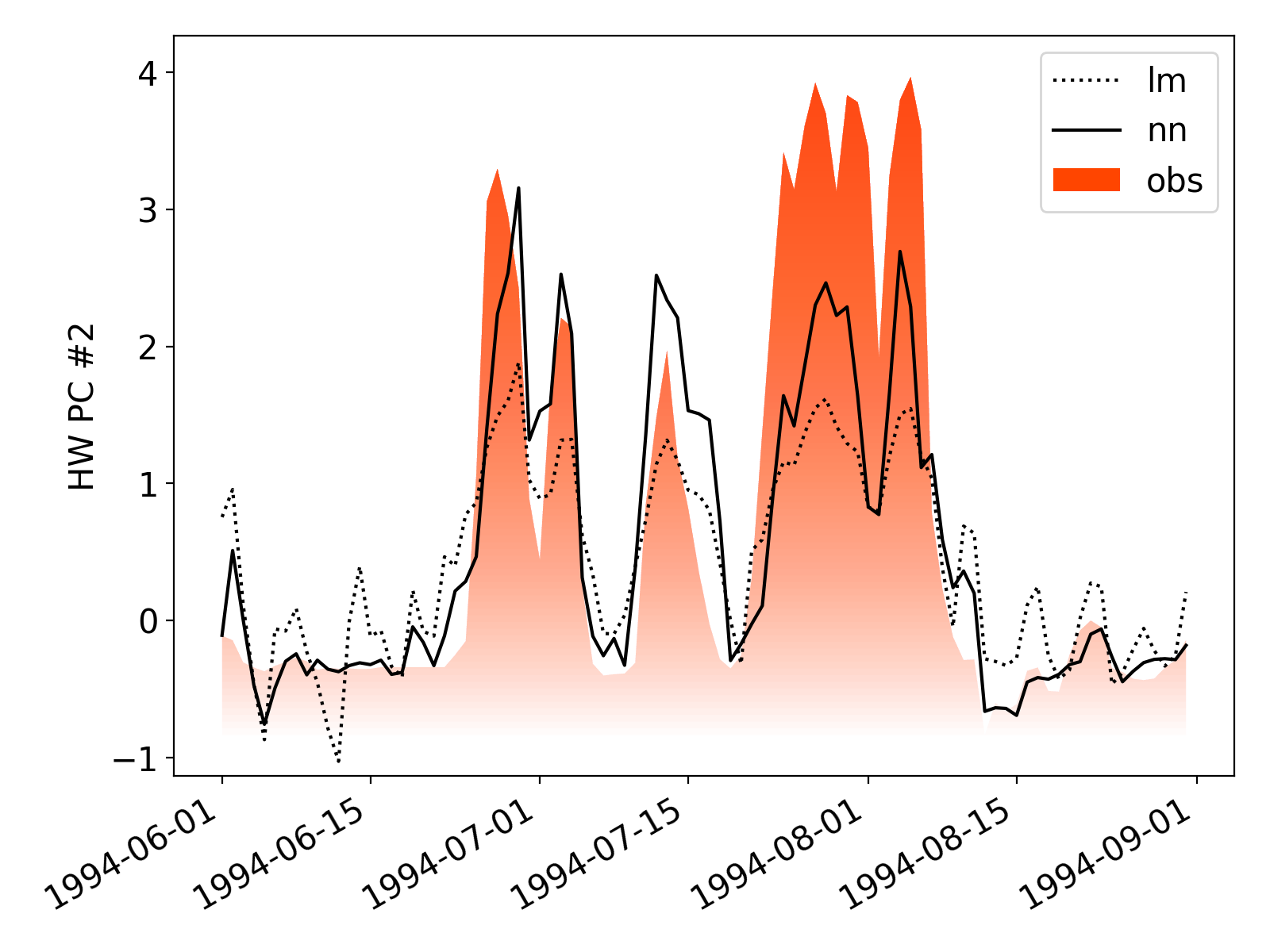}
    \caption{Observed and modelled HW principal component number two (Central EU) in summer of 1994.}
    \label{fig:ex2019}
\end{figure}

\begin{table}
    \centering
    \begin{tabular}{llrr}
    {} &      region &    nn &    lm \\\hline
       &       total &  0.70 &  0.43 \\
    1  &      Baltic &  0.65 &  0.38 \\
    2  &  Central EU &  0.72 &  0.41 \\
    3  &      Russia &  0.66 &  0.36 \\
    4  &    Anatolia &  0.67 &  0.36 \\
    5  &  North Cape &  0.74 &  0.52 \\
    6  &      Iberia &  0.72 &  0.44 \\
    7  &      Poland &  0.65 &  0.37 \\
    8  &      Balkan &  0.74 &  0.45 \\
    9  &  North East &  0.66 &  0.40 \\
    10 &   Skagerrak &  0.74 &  0.58 \\
    \end{tabular}
    \caption{Total $R^2$ and $R^2$ of the individual logistic principal components for the non-linear (nn) and linear model (lm).}
    \label{tab:R2local}\end{table}

Using 20 principal components from each of the input variables listed in table \ref{tab:predictors}, our linear regression model achieves $R^2\approx 0.43$, while the non-linear model reaches $R^2\approx 0.70$ (see table \ref{tab:R2local}). To understand this remarkable difference, it is instructive to consider an example prediction. Figure \ref{fig:ex2019} shows both the observed and the two predicted time series for the second principal heat wave component throughout the 1994 summer season. We observe that the central EU region was affected by three separate heat wave events, represented by prominent positive peaks in the corresponding PC. In early June and late August, the observed values remain nearly constant at a slightly negative level.

Our non-linear network model is able to reproduce both the nearly constant interval and the large peaks reasonably well. The linear model, on the other hand, loosely follows the same evolution of events, but is unable to trace the extreme peaks or remain constant for an extended period of time. This inability to reproduce the intermittent behaviour of the heat waves explains the greatly decreased $R^2$. Our non-linear network, on the other hand, contains a ReLU activation which precisely mirrors the threshold in our heatwave definition.\\

Table \ref{tab:R2local} also lists the values of $R^2$ for each logistic principal component individually. Our non-linear model appears to perform similarly well across all regions while the performance of the linear model is somewhat more variable. Most notably, the Scandinavian regions 5 and especially 10 are represented  substantially better than any other part of the domain.

\subsection{Explaining the model performance}

\begin{figure}
    \centering
    \includegraphics[width=.66\textwidth]{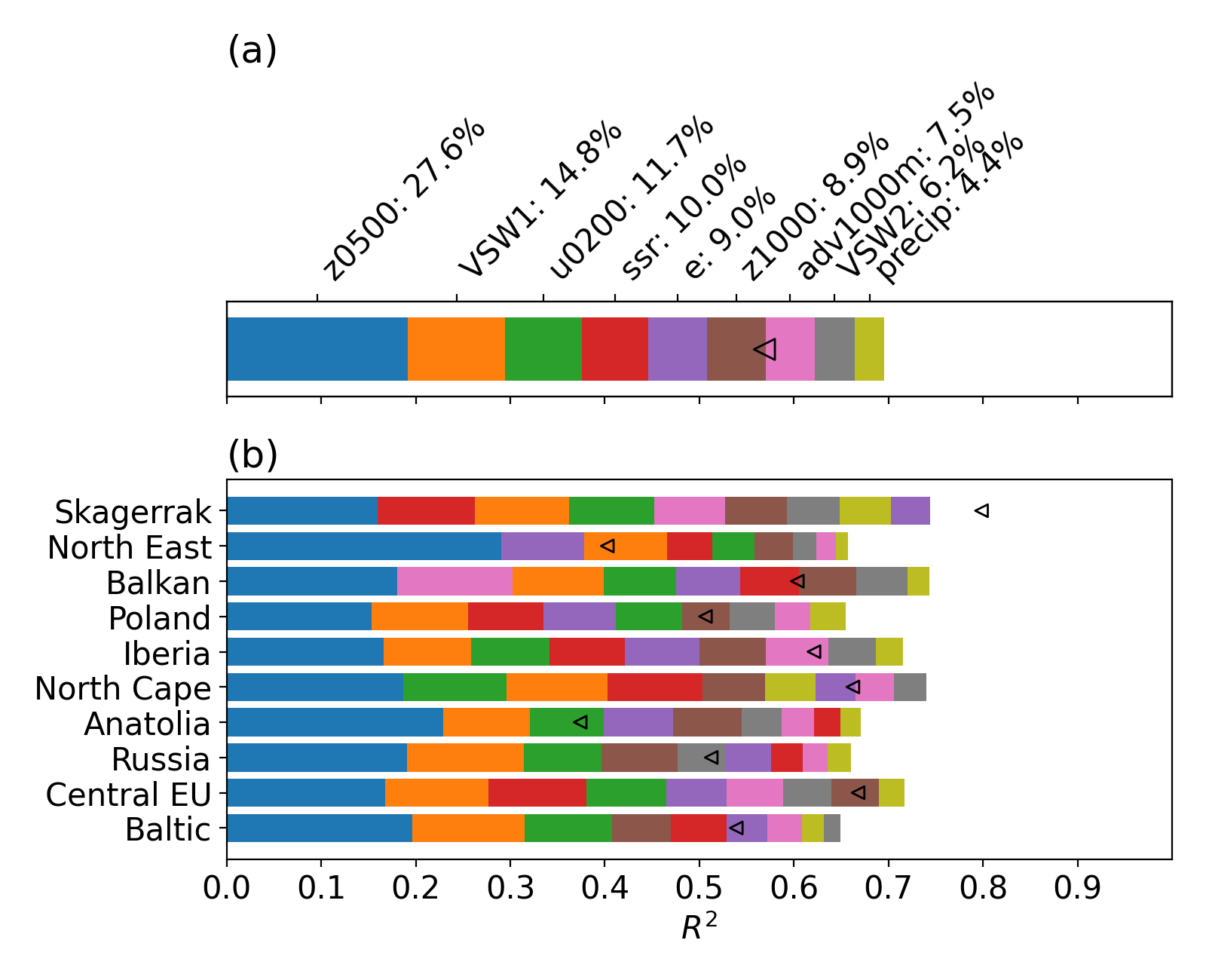}
    \caption{Decomposition of the overall (a) and regional (b) $R^2$ via Shapley values. Each stack is ordered from largest to smallest contributions, colors follow the order of panel (a). Triangles mark the value of the total (negative) interaction (sum over the off-diagonal elements $\varphi_{i,j\neq i}$).}
    \label{fig:shap}
\end{figure}

To decompose the roughly $70\,\%$ explained variance of our full neural network model into contributions from each of the nine predictor variables, we train and evaluate each of the 511 other possible models. From the differences between all 512 values of $R^2$, we compute the exact Shapley values according to equation \ref{eq:shapley}. The resulting decomposition is shown in figure \ref{fig:shap}\,a. With a contribution of $28\,\%$, 500\,hPa geopotential emerges as the top contributor, followed by top level soil moisture with almost 15\,\%. Most other variables make similar, minor contributions between 7 and 10\,\%, second level soil moisture and precipitation are least informative.\par
Figure \ref{fig:shap}\,(b) displays the analogous decompositions for the individual $R^2$ of each principal component. We find that z500 is not only the best predictor overall but also in each individual region, while precipitation and VSW2 are always among the weakest contributors. The influence of other variables, however, can vary significantly from region to region: ssr, for example, is among the top three variables for the central EU, Poland and Skagreak components; in other regions such as Russia and Anatolia, ssr is close to tied with precipitation as the worst predictor. Near surface advection is among the weakest predictors in all regions except the Balkans, where it is actually in second place. 

\begin{figure}
    \centering
    \includegraphics[width=.66\textwidth]{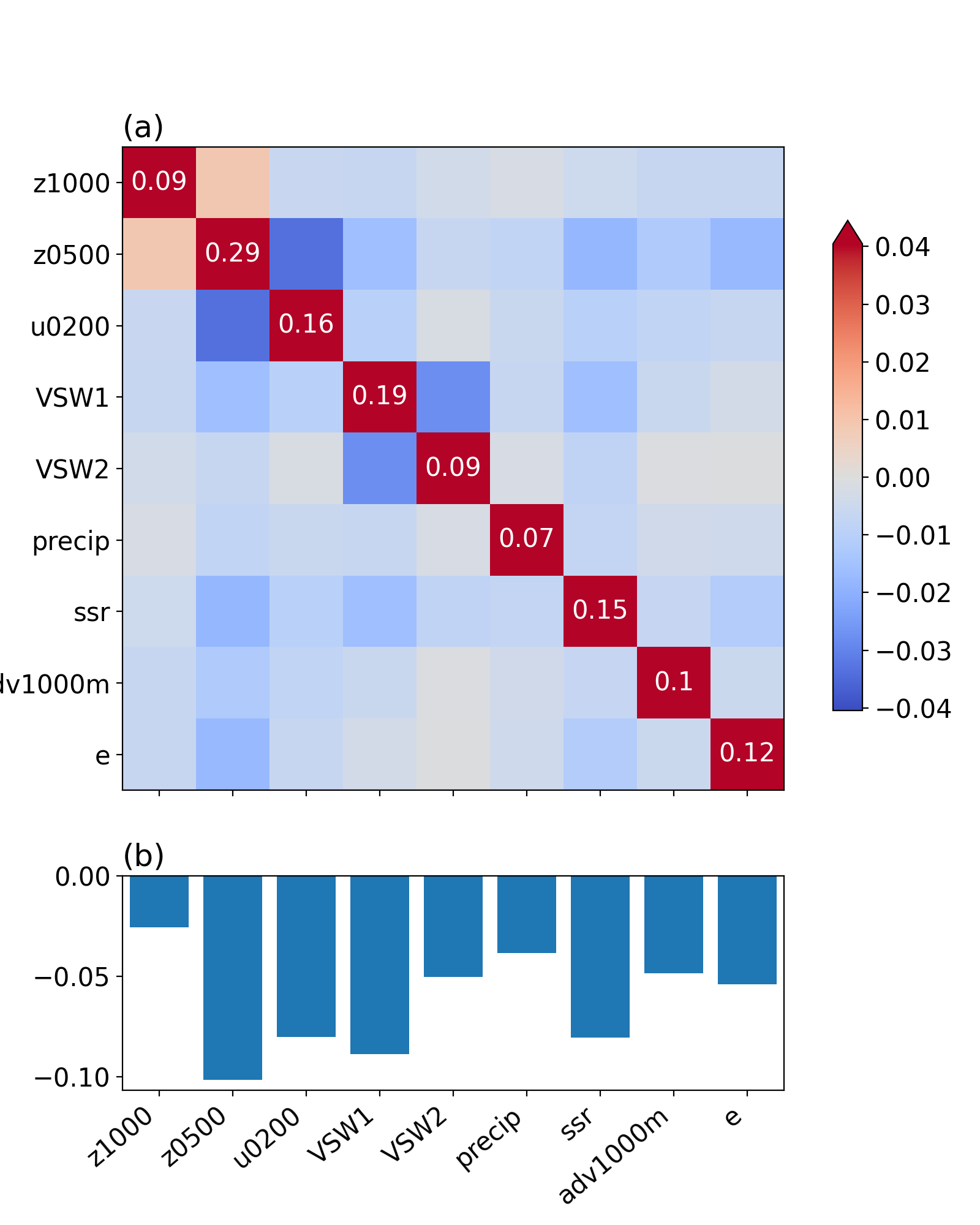}
    \caption{(a) matrix of Shapley interaction values (equation \ref{eq:hausken1}) for the overall $R^2$ of the neural network model. The values on the main diagonal are given as white numbers. (b) sum of all interactions (off-diagonal elements) for each variable.}
    \label{fig:hausken}
\end{figure}

Next, the Shapley values from figure \ref{fig:shap}\,(a) are further decomposed into Shapley interactions according to equation \ref{eq:hausken1}. The resulting matrix, shown in figure \ref{fig:hausken}, has predominantly negative entries on the off-diagonals, indicating a wide-spread overlap of information between the various predictors. Especially the two upper-level flow features (z500, u200) and the two soil moisture levels (VSW1, 2) have strongly negative interactions: if one of the two is already part of the model, the other one has little added value. Surprisingly, z500 and z1000, the two variables with the most strongly correlated PCs (not shown), are the only pair with a net positive interaction, meaning that one is more useful in the presence of the other. As a result, the total interaction (figure \ref{fig:hausken}\,b), which is negative for all variables, is weakest for z1000. Conversely, the best predictors z500 and VSW1 also have the strongest negative interactions, their values are reduced by a third and almost half, respectively.\par 

An analogous interaction matrix can be computed for each of the ten heatwave PCs individually, enabling a further detailed analysis of regional differences. Here, we merely summarize this information by the total interaction strength (sum over all off-diagonal interaction elements), shown as triangular markers in figure \ref{fig:shap}. Compared to the model performance itself, the interactions vary more strongly across the domain, while higher $R^2$ is generally correlated with greater overlap between variables. This is particularly striking for the Skagerrak region, where the interaction is greater than $R^2$ itself and the distribution among variables is almost uniform. The situation is similar for the North Cape, these two also being the only components with a negative interaction between z500 and z1000 (not shown). Overall, the apparently simpler nature of the modelling problem may in part explain why these two components are exceptionally well modelled by the linear model (cf. table \ref{tab:R2local}).

\subsection{Explaining trends}

\begin{figure*}
    \centering
    \includegraphics[width=\textwidth]{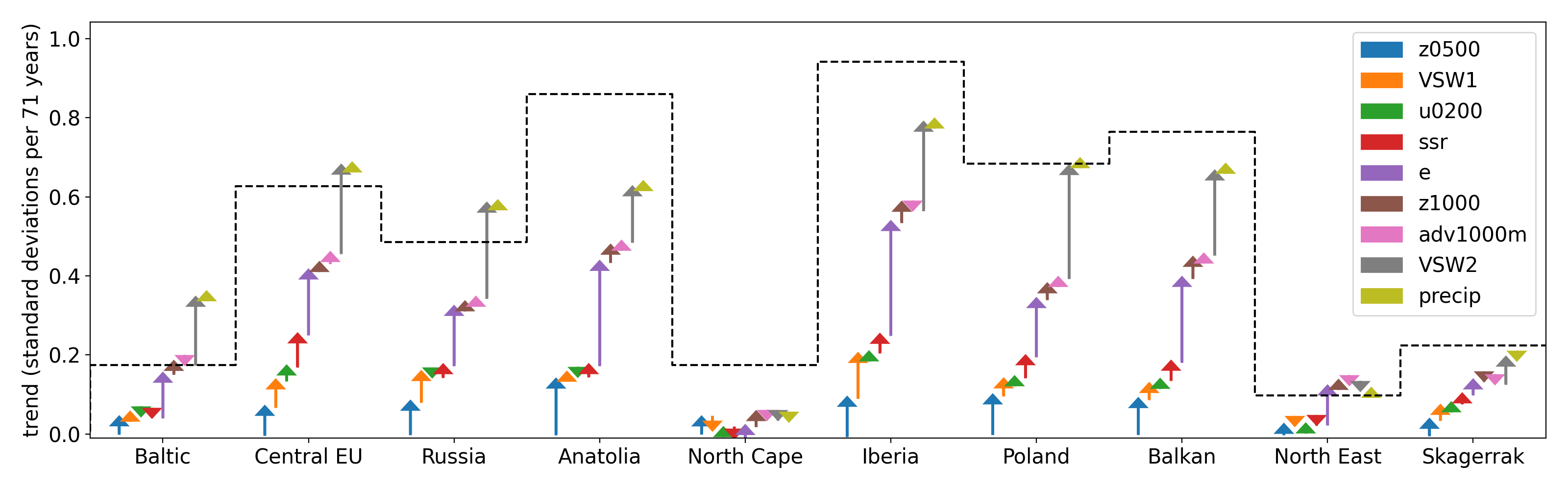}
    \caption{Observed (dashed line) and modelled trend (arrows) for the principal heat wave components. The length and direction of the arrows indicate the contribution of each predictor to the modelled trend, computed via Shapley values.}
    \label{fig:trendshap}
\end{figure*}

So far, we have limited our analysis to decompositions of the model performance, measured by $R^2$. In principle, the general definition of Shapley values can be applied to \textit{any} characteristic of the model. As an example, we replace $f(S)$ in equation \ref{eq:shapley} by the modeled linear trend of one of the heatwave PCs. Trends are computed as the slopes of a linear regression against a sequence from zero (1.6.1950) to one (31.8.2020). Since the components were scaled to unit variance, the resulting observed trend values, shown as a dashed line in figure \ref{fig:trendshap}, indicate changes of up to one standard deviation over the course of the 71 year time series. Here, we find pronounced regional differences, with almost no observed trend in the Northern regions (Baltic, Skagerrak, North Cape and North East) and pronounced changes especially in Iberia, Anatolia and the Balkans. This overall pattern is reproduced by the neural network model, which underestimates the most extreme trends while slightly overestimating the changes in the Baltic and Russian region. The contributions of the individual predictors are visualized as arrows instead of stacked bars as in figure \ref{fig:shap} to allow for positive and negative contributions. We see that most of the information on trends can be extracted from evaporation and second level soil moisture, two variables with comparatively little impact on the model performance. The relative importance of the two varies from region to region with soil moisture dominating in Poland and evaporation in Anatolia. In some regions, a modest part of the trend can also be learned from z500.

\section{Discussion}\label{sec:disc}
Looking at the decomposition in figure \ref{fig:shap}\,(a) one might be tempted to draw conclusions like ``27.6\,\% of European heatwaves are caused the upper level dynamics''. To avoid such erroneous interpretations, we must reflect carefully on what exactly the Shapley values, used in the way we have done here, mean. Recall that we have set $f(S)$ in the game theoretic definition (equation \ref{eq:shapley}) to the success of a model that reproduces heat waves from the variables contained in $S$. The values $\varphi_i$ thus tell us how much of that success we should attribute to each variable. If we assume that the model makes optimal use of the information it is presented with, we can interpret $\varphi_i$ as an amount of information on heatwaves that can be gleamed from variable $i$. It is important to note that this is only a measure of (non-linear) association, not \textit{causation}: whether or not one event causes another cannot be decided by purely descriptive statistics.
The results of our analysis are then similar to individual correlations or the coefficients of a multiple linear regression (as in \cite{suarez2020}), but, crucially, allow for nonlinear relationships as well as overlap and interactions between different predictors. 

For example, positive interactions can render uni-variate linear correlations completely unhelpful when some variables only ``reveal'' their information in combination with others. In our framework, we can not only reward such cases appropriately, but also directly quantify interactions via the interaction matrix. One interesting example here is the positive synergy between z500 and z1000. Had we attempted to build the best heatwave model possible by conventional means, we would likely have dropped z1000 from the set of predictors (either manually or via a variable selection approach like LASSO) due to the strong correlation with z500 and comparatively weaker correlation with heatwaves.
Using Shapley values, however, we have found that, instead of overlapping, these two predictors actually improve each other -- they are highly correlated across the whole data set, but have complementary behaviour in relationship to heat waves. This makes sense from a physical perspective: much of our mid-latitude weather is driven by dynamical pressure systems with interacting  anomalies of the same sign in upper and lower levels. Persistent temperature extremes, however, lead to the formation of thermal pressure systems with opposite anomalies near the surface, in our case heat lows. The fact that the signs of the upper and lower anomaly differ, can be used to distinguish a heatwave situation from regular, moving pressure systems. This explanation is supported by the fact that the interaction between the two geopotentials is weakest in the Northernmost components 5, 9 and 10 (not shown), where the absolute temperatures are low.  

An important methodological distinction between our approach and that of \cite{suarez2020}, as well as \cite{felsche2021} and \cite{vanStraaten2022}, is that those authors explain the impact of individual inputs on an already trained model. Linear models as used by \cite{suarez2020} are self explanatory because their coefficients can be inspected directly to understand the model behaviour. The behaviour of a trained non-linear model (or any other function) can be explained in terms of Shapley values as well. We cannot technically ``withhold'' an input from such a model, but the absence of features can be simulated, for example by randomly permuting their values in time. This idea, popularized with great success by \cite{lundberg2017}, and used by \cite{felsche2021} and \cite{vanStraaten2022}, makes sense when we intend to use the model for some purpose, for example to issue seasonal predictions. In this setting, an explanation of the actual model's behaviour helps users asses its trustworthiness and identify reasons for its shortcomings \citep{ribeiro2017}. Since we care only about the information content of the predictors, and not about the model itself, retraining the models is more appropriate. In particular, the approach of \cite{lundberg2017} assumes independence between features which can lead to misleading results \citep{aas2021}: a single model can easily decide on one out of several correlated features, giving little or no weight to the others.   

\begin{figure}
    \centering
    \includegraphics[width=.66\textwidth]{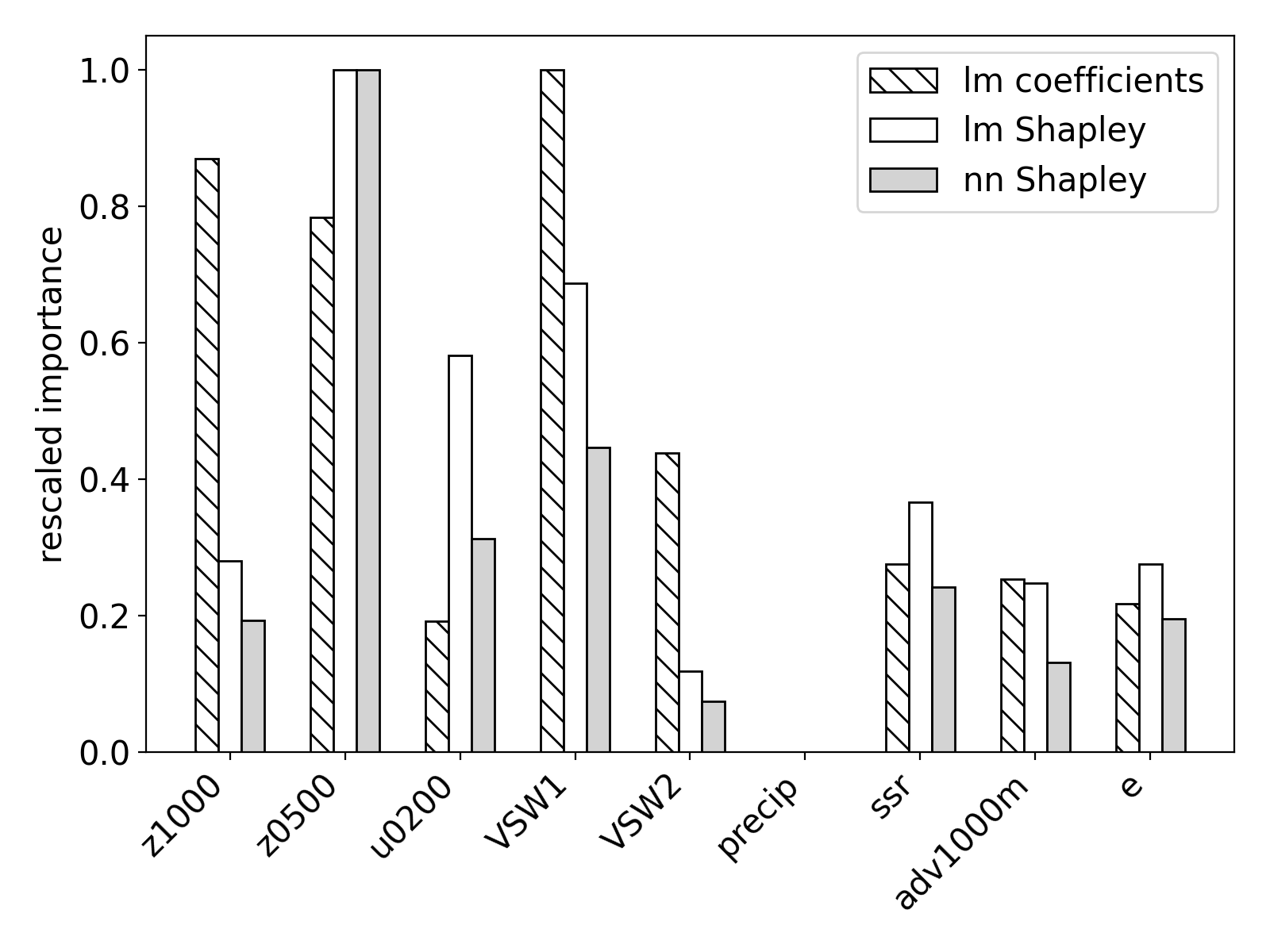}
    \caption{Mean absolute regression coefficients and Shapley values for the linear model and Shapley values for the non-linear model, all re-scaled to the range $[0,1]$ for comparison. Non-linear Shapley values are the same as in figure \ref{fig:shap}(a), those of the linear model were obtained by the same procedure of re-training all possible models.}
    \label{fig:lmcoef}
\end{figure}

For a simple comparison of the different kinds of explanations, we can return to the linear model used in section \ref{sub:regr}. Figure \ref{fig:lmcoef} shows the absolute regression coefficients, averaged over all regions and cross validation folds. From this analysis, we would conclude that soil moisture is most closely connected to heat waves, followed by z1000 before z500. This result differs substantially not only from the $\varphi_i$ of the non-linear model, but also from those of the linear model, both of which are included in figure \ref{fig:lmcoef}. Indeed, despite the big gap in performance, the two models agree closely on the ranking of the predictors, with the linear model gaining slightly less information from z500.

We conclude this discussion with a number of caveats and limitations that must be kept in mind when interpreting our results. 
Firstly, all outcomes are conditional on the model used to represent the relationships. If the complexity of the model is insufficient, we cannot give due credit to variables whose impact on heatwaves was not learned. Conversely, a model that is too complex to be adequately trained on the available data will over-fit. In this case, predictors can even have negative impact on model performance, i.e., $\varphi_i<0$, by ``distracting'' from the relevant relationships, thus contradicting our interpretation in terms of information that can be learned. We have found a reasonable compromise in our small neural network with dropout layers and early stopping to combat over-fitting. 
Secondly, the Shapley value of each variable is not an absolute measure of its connection to heat waves, since it necessarily depends on the overall selection of included variables.\par

\section{Conclusion}\label{sec:outro}

It is generally accepted that the occurrence and duration of heatwaves are determined by an interplay of different processes including upper and lower level dynamics, radiation and surface feedbacks. In a very specific sense, discussed in detail in the previous section, we have been able to put numbers on the amount of information on heatwaves present in various variables related to the different processes. The result that z500 and upper level soil moisture are most characteristic of heat waves is hardly surprising. Other details, such as the comparatively minor role of advection and radiation or the more than 2-fold dominance of z500 over jet-level wind speeds, were less clear a priori. With the help of interaction values, we have further cleared up, which variables overlap strongly (for example z500 and u200) and which even have positive synergy (z500 and z1000). 
To achieve these results, we have employed and slightly adapted two non-standard methods to our goal of explaining climate extremes: the logistic PCA of \cite{landgraf2020} and Shapley values and interactions \citep{lipovetsky2001,hausken2001}. Both of these techniques offer interesting directions for future research.\par

The lPCA offers a new way of compactly representing fields of binary events. Compared to clustering, as used by \cite{stefanon2012} and \cite{vanStraaten2022} for heatwaves, the reduction is less drastic (n real numbers instead of a single integer) and the original field can be reconstructed if needed. The technique is applicable to any binary event, regardless  and how simple or complex the definition and which or how many variables it involves. For example, lPCA could be used to compactly compare different heat wave definitions or encode the occurrence of cold spells, jet streaks or atmospheric rivers, as well as compound climate events \citep{zscheischler2020}. The formulation as a neural network simplifies the explanation, speeds up the implementation on large scales and renders the lPCA more flexible. Here, we have simply added an $L_1$ penalty to obtain sparse basis functions that are easier to interpret. For some purposes, an even more efficient compression via non-linear auto-encoders would be an interesting next step.\par

We have seen that Shapley values emerge very naturally when we attempt to distribute a model's skill among the input variables. In the absence of independent predictors, this approach is recommended not only for non-linear neural networks but also for simpler models. The computation requires minimal dedicated software and can, at least to some extent, be approximated by sampling (see appendix \ref{app:sample}). Great unused potential lies in the generality of equation \ref{eq:shapley}: in principle we can decompose the output of any algorithm, no matter how simple or complex, as long as we can evaluate it for any subset of its inputs. When withholding an input is not technically possible, it can be simulated following \cite{lundberg2017}. One example is the decomposition of other model attributes besides the skill, as we have demonstrated for the linear trend of the heatwave PCs. Beyond this, however, the world of earth system science is full of complex non-linear functions waiting to be explained.


\appendix
\section{Approximate Shapley values by sampling}\label{app:sample}

\begin{figure}
    \centering
    \includegraphics[width=.66\textwidth]{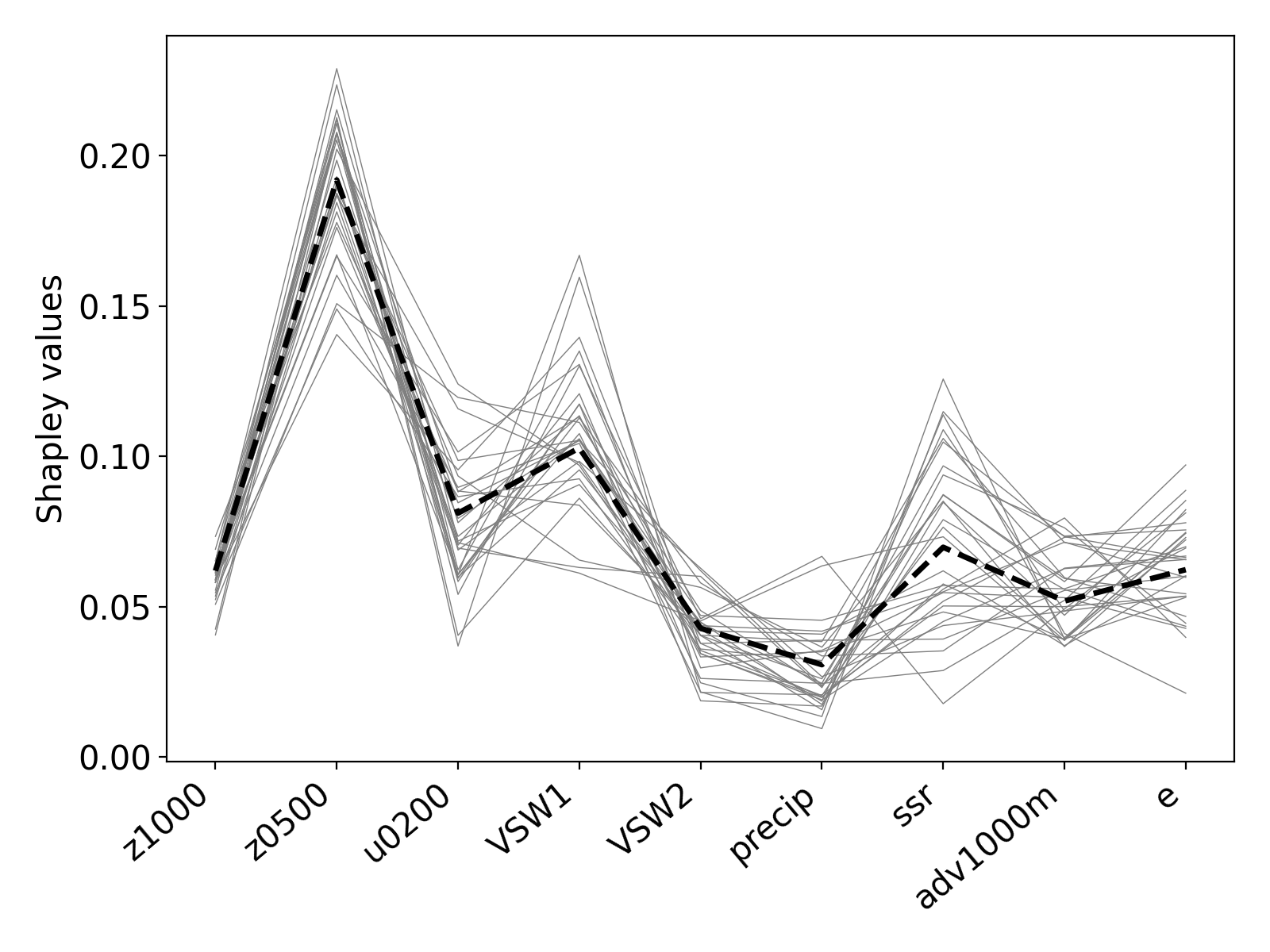}
    \caption{Exact Shapley values as in figure \ref{fig:shap}\,a (dashed line) and 50 realizations of sampled Shapley values, each based on 20 repetitions of algorithm \ref{alg:randomshap} (grey).}
    \label{fig:sampleshap}
\end{figure}

In section \ref{sec:res} we have discussed the exact Shapley values computed from all 512 possible models. Since the number of possibilities doubles with each additional variable, the exact approach eventually becomes intractable. We can obtain approximate results by averaging over several repetitions of algorithm \ref{alg:randomshap}. Figure \ref{fig:sampleshap} reveals that 20 repetitions, necessitating the estimation of around 122 out of 512 models, typically give a reasonably close approximation of the exact result: almost all sampled curves show a correlation greater than 0.8 and RMSE below 0.05 with respect to the true values. It should be noted that the number of necessary repetitions will depend on the specific problem at hand. In our case, 20 repetitions are more than sufficient to estimate the value for z1000, which has only a small net interaction. More strongly interacting variables have higher uncertainty and require a larger sample size to arrive at a confident estimate.

\bibliography{save}

\end{document}